\documentclass{ws-procs9x6}
\usepackage{graphicx}
\usepackage{latexsym}
\usepackage{amssymb}

\newcommand{\dsp}{\displaystyle}
\newcommand{\ba}{\begin{eqnarray}}
\newcommand{\ea}{\end{eqnarray}}
\newcommand{\be}{\begin{equation}}
\newcommand{\ee}{\end{equation}}
\newcommand{\re}{\mbox{Re}}
\newcommand{\im}{\mbox{Im}}
\newcommand{\kob}{{\overline{K^0}}}
\newcommand{\epspeps}{\ensuremath{\varepsilon^\prime/\varepsilon}}

\newcommand{\hepph}[1]{{hep-ph/#1}}

\newcommand{\citebk}[1]{[\hspace{0.mm}\raisebox{-1.85mm}[0mm][0mm]
  {\Large\cite{#1}}\hspace{-0.1mm}]}

\newcommand{\citebkcap}[1]{[\hspace{0.mm}\raisebox{-1.5mm}[0mm][0mm]
  {\large\cite{#1}}\hspace{-0.2mm}]}

\begin{document}
\thispagestyle{empty}
\begin{flushright}
LU TP 02-27\\
hep-ph/0207082\\
July 2002
\end{flushright}
\vfill
\begin{center}
{\huge{\bf  Penguins 2002: Penguins in $K\to\pi\pi$ Decays}$^a$}
\end{center}

\vfill

\begin{center}
{\large\bf Johan~Bijnens}\\[1cm]
{\large Department of Theoretical Physics, Lund University,\\[0.5cm]
S\"olvegatan 14A, S22362 Lund, Sweden}
\end{center}

\vfill
\begin{center}
{\bf Abstract}
\end{center}
{This talk contains a short overview of the history of the interplay of
the weak and the strong interaction and $CP$-violation. It describes
the phenomenology and the basic physics mechanisms involved in the
Standard Model calculations of $K\to\pi\pi$ decays with an emphasis
on the evaluation of Penguin operator matrix-elements.}

\vfill

\noindent $^a$
{\small
Presented at the
Symposium and Workshop "Continuous Advances in QCD 2002/Arkadyfest,"
Minneapolis, USA, 17-23 May 2002, to be published in the proceedings.}

\setcounter{page}{0}
\newpage

\title{Penguins 2002: Penguins in $K\to\pi\pi$ Decays}

\author{Johan Bijnens}

\address{Department of Theoretical Physics 2, Lund University,\\
S\"olvegatan 14A, S22362 Lund, Sweden\\
E-mail: bijnens@thep.lu.se}


\maketitle

\abstracts{
This talk contains a short overview of the history of the interplay of
the weak and the strong interaction and $CP$-violation. It describes
the phenomenology and the basic physics mechanisms involved in the
Standard Model calculations of $K\to\pi\pi$ decays with an emphasis
on the evaluation of Penguin operator matrix-elements.}

\section{Introduction}

\begin{figure}
\begin{center}
\includegraphics[width=0.455\textwidth]{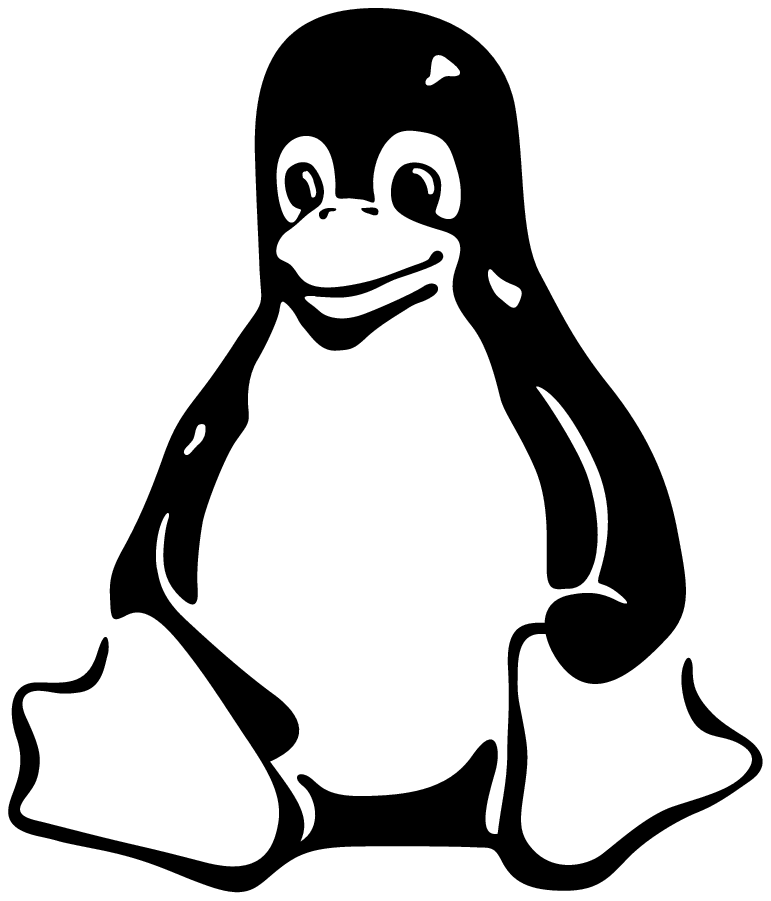}
\nolinebreak
\includegraphics[width=0.379\textwidth]{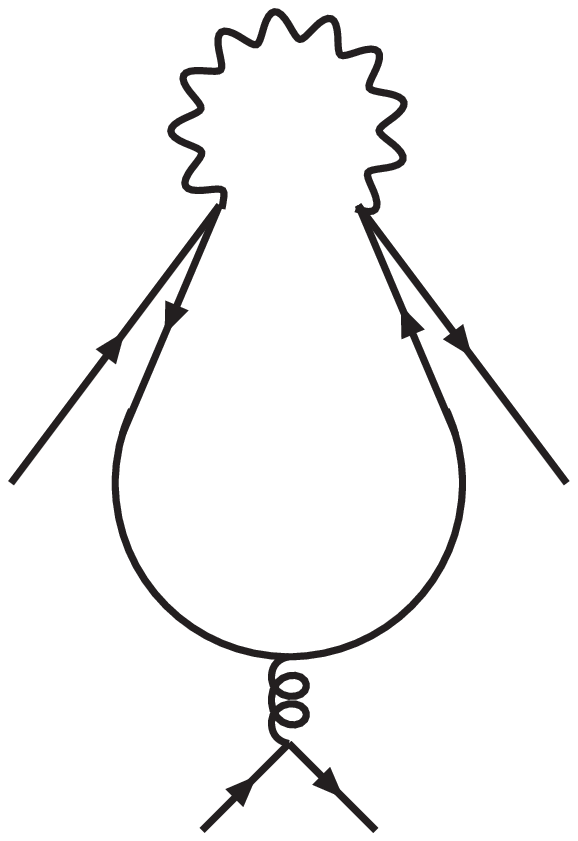}
\end{center}
\caption{\label{fig1} The comparison between a (Linux) Penguin and
the Penguin diagram. (Linux Penguin from Neal Tucker,
({\tt http://www.isc.tamu.edu/$\sim$lewing/linux/}).)}
\end{figure}

In this conference in honour of Arkady Vainshtein's 60th birthday, a
discussion of the present state of the art in analytical calculations
of relevance for Kaon decays is very appropriate given Arkady's large
contributions to the field. He has summarised his own contributions
on the occasion of accepting the Sakurai Prize.\cite{Penguin3}
This also contains the story of how Penguins, at least the diagram
variety, got their name. In Fig.~\ref{fig1} I show what a real
(Linux) Penguin looks like and the diagram in a ``Penguinized'' version.
This talk could easily have had other titles, examples are
``QCD and Weak Interactions of Light Quarks'' or
``Penguins and Other Graphs.'' In fact I have left out many
manifestations of Penguin diagrams. In particular I do not cover the
importance in $B$ decays where Penguins were first experimentally verified
via $B\to K^{(*)}\gamma$, but are at present more considered a nuisance
and often referred to as ``Penguin Pollution.'' Penguins also play
a major role in other Kaon decays, reviews of rare decays where
pointers to the literature can be found are
Refs~\citebk{Isidori,Littenberg,Buras2}.

In Sect.~\ref{history} I give a very short historical overview.
Sects~\ref{deltaIhalf} and \ref{epspeps} discuss the main physics
issues and present the relevant phenomenology of the $\Delta I=1/2$ rule
and the $CP$-violating quantities $\varepsilon$ and $\epspeps$.
The underlying Standard Model diagrams responsible for $CP$-violation are 
shown there as well. The more challenging part is to actually evaluate
these diagrams in the presence of the strong interaction. We can
distinguish several regimes of momenta which have to be treated using
different methods. An overview is given in Sect.~\ref{EFT}
where also the short-distance part is discussed. The more difficult
long-distance part has a long history and some approaches are mentioned,
but only my favourite method, the $X$-boson or fictitious Higgs exchange,
is described in more detail in Sect.~\ref{Xboson} where I also present results
for the main quantities. For two particular matrix-elements, 
those of the electroweak
Penguins $Q_7$ and $Q_8$, a dispersive analysis allows to evaluate these
in the chiral limit from experimental data. This is discussed in
Sect.~\ref{dispersive}. We summarise our results in the conclusions and compare
with the original hopes from Arkady and his collaborators.
This talk is to a large extent a shorter version of the
review~\citebk{QCDweak}.

\section{A short historical overview}
\label{history}

The weak interaction was discovered in 1896 by Becquerel when he discovered
spontaneous radioactivity. The next step towards a more fundamental
study of the weak interaction was taken in the 1930s when the neutron was
discovered and its $\beta$-decay studied in detail. 
The fact that the proton and electron energies did not add up
to the total energy corresponding to the mass of the neutron,
made Pauli suggest the neutrino as a solution.
Fermi then incorporated it in the first full fledged theory of the
weak interaction, the famous Fermi four-fermion~\cite{Fermi} interaction. 

\be
\label{Fermi1}
{\mathcal L}_{\mbox{Fermi}} =
\frac{G_F}{\sqrt{2}}\; [\overline p \gamma_\mu\left(1-\gamma_5\right) n]\;
[\overline e \gamma^\mu\left(1-\gamma_5\right)\nu]\,.
\ee

The first fully nonhadronic weak interaction came after world-war two
with the muon discovery and the study of its $\beta$-decay.
The analogous Lagrangian to Eq.~(\ref{Fermi1}) was soon written down.
At that point T.D.~Lee and C.N.~Yang~\cite{LeeYang} realized that there was no
evidence that parity was conserved in the weak interaction. This quickly
led to a search for parity violation both in nuclear decays~\cite{CSWu}
and in the decay chain
$\pi^+\to\mu^+\nu_\mu\to e^+\nu_e\bar\nu_\mu\nu_\mu$.\cite{FriedmanTelegdi}
Parity violation
was duly observed in both cases. These experiments and others
led to the final form of the
Fermi Lagrangian given in
Eq.~(\ref{Fermi1}).\cite{SudarshanMarshak,FeynmanGell-Mann}

During the 1950s steadily more particles were discovered
providing many puzzles.
These were solved by the introduction of
strangeness,\cite{Pais,Gell-Mann}
of what is now known as the $K_L$ and the $K_S$~\cite{Gell-MannPais} and
the ``eightfold way'' of classifying the hadrons
into symmetry-multiplets.\cite{eightfold}

Subsequently Cabibbo realized that the weak interactions of
the strange particles were very similar to those of the
nonstrange particles.\cite{Cabibbo} He proposed that the weak interactions
of hadrons occurred through a current which was a mixture of the
strange and non-strange currents with a mixing angle now universally
known as the Cabibbo angle. The hadron symmetry group
led to the introduction of quarks~\cite{quarks} as a means of
organising which $SU(3)_V$ multiplets were present in the
spectrum.

In the same time period the Kaons provided another surprise.
Measurements at
Brookhaven~\cite{CCFT} indicated that the long-lived state, the $K_L$,
did occasionally decay to two pions in the final state as well,
showing that $CP$ was violated.
Since the $CP$-violation was small, explanations could be sought at
many scales, an early phenomenological analysis can be found
in Ref. ~\citebk{WuYang}, but as 
the socalled superweak model~\cite{Wolfenstein} showed,
the scale of the interaction involved in $CP$-violation could be much higher.

The standard model for the weak and electromagnetic interactions of leptons
was introduced in the same period.
The Fermi theory is nonrenormalisable.
Alternatives based on Yang-Mills~\cite{Yang-Mills}
theories had been proposed by Glashow~\cite{Glashow} but struggled with
the problem of having massless gauge bosons. This was solved by 
the introduction of the
Higgs mechanism by Weinberg and Salam. The model
could be extended to include the weak interactions of hadrons by adding
quarks in doublets, similar to the way the leptons were included.
One problem this produced was that loop-diagrams provided a much too
high probability for the decay $K_L\to\mu^+\mu^-$ compared to
the experimental limits. These socalled flavour changing neutral currents
(FCNC) needed to be suppressed. The solution was found in
the Glashow-Iliopoulos-Maiani mechanism.\cite{GIM} A fourth quark, the charm
quark, was introduced beyond the up, down and strange quarks.
If all the
quark masses were equal, the dangerous loop contributions to FCNC processes
cancel, the socalled GIM mechanism.
This allowed a prediction of the charm quark
mass,~\cite{GaillardLee1} soon confirmed with
the discovery of the $J/\psi$.

In the mean time, 
QCD was formulated.\cite{QCD} The property of
asymptotic freedom~\cite{Asympt} was established which explained why
quarks at short distances could behave as free particles and at the
same time at large distances be confined inside hadrons.

The study of Kaon decays still went on, and an already old problem,
the $\Delta I=1/2$ rule saw the first signs of a solution. It was
shown~\cite{GaillardLee2,AltarelliMaiani} that the short-distance QCD part
of the nonleptonic weak decays provided already an enhancement of
the $\Delta I=1/2$ weak $\Delta S=1$ transition over the $\Delta I=3/2$
one. The ITEP group extended first the Gaillard-Lee analysis
for the charm mass,\cite{Vainshtein1} but then realized that in
addition to the effects that were included in
Refs~\citebk{GaillardLee2,AltarelliMaiani},
there was a new class of diagrams that only contributed to
the $\Delta I=1/2$ transition.\cite{Penguin1,Penguin2} While, as we
will discuss in more detail later, the general class of these
contributions, the socalled Penguin-diagrams, is the most likely main
cause of the $\Delta I=1/2$ rule, the short-distance part of them
provide only a small enhancement contrary to the original hope.
A description of the early history of Penguin diagrams, including the
origin of the name, can be found in the 1999 Sakurai Prize lecture  of
Vainshtein.\cite{Penguin3}

Penguin diagrams at short distances provide nevertheless a large amount
of physics. The origin of $CP$-violation was (and partly is)
still a mystery. The superweak model explained it, but introduced new
physics that had no other predictions. 
Kobayashi and Maskawa~\cite{KobayashiMaskawa} realized that
the framework established by Ref.~\citebk{GIM} could be extended
to three generations.
The really new aspect this brings in is that $CP$-violation
could easily be produced at the weak scale and not at the much
higher superweak scale. In this Cabibbo-Kobayashi-Maskawa (CKM)
scenario, $CP$-violation comes from
the mixed quark-Higgs sector, the Yukawa sector,
and is linked with the masses and mixings of the quarks.
Other mechanisms at the weak scale
also exist, as e.g. an extended Higgs sector.\cite{WeinbergCP}

The inclusion of the CKM mechanism into the
calculations for weak decays was done by Gilman and
Wise~\cite{GilmanWise1,GilmanWise2} which provided the prediction
that $\epspeps$ should be nonzero and of the order of $10^{-3}$.
Guberina and Peccei~\cite{GuberinaPeccei} confirmed this.
This prediction spurred on the experimentalists and after
two generations of major experiments, NA48 at CERN and KTeV at Fermilab
have now determined this quantity and the qualitative prediction
that $CP$-violation at the weak scale exists is now confirmed.
Much stricter tests of this picture will happen at other Kaon
experiments as well as in $B$ meson studies.

The $K^0$-$\kob$ mixing has QCD corrections and $CP$-violating contributions
as well. The calculations of these required a proper treatment
of box diagrams and inclusions of the effects of the $\Delta S=1$
interaction squared. This was accomplished at one-loop
by Gilman and Wise a few years later.\cite{GilmanWise3,GilmanWise4}

That Penguins had more surprises in store was shown some years later
when it was realized that the enhancement originally expected on chiral grounds
for the Penguin diagrams~\cite{Penguin1,Penguin2} was present,
not for the Penguin diagrams with gluonic intermediate states, but
for those with a photon.\cite{BijnensWise} This contribution
was also enhanced in its effects by the $\Delta I=1/2$ rule. 
This lowered the
expectation for $\epspeps$, but it became significant
after it was found that the top quark had a very large mass.
Flynn and Randall~\cite{RandallFlynn} reanalysed the electromagnetic
Penguin with a large top quark mass and included also $Z^0$ exchange.
The final effect was that
the now rebaptized electroweak Penguins could have a very large contribution
that could even cancel the contribution to $\epspeps$ from
gluonic Penguins. This story still continues at present and the
cancellation, though not complete, is one of the major impediments
to accurate theoretical predictions of $\epspeps$.

The first calculation of two-loop effects in the short-distance part
was done in Rome~\cite{twoloopold} in 1981.
The value of $\Lambda_{\mbox{QCD}}$ has risen
from values of about 100~MeV to more than 300~MeV. A full calculation
of all operators at two loops thus became necessary, taking into account
all complexities of higher order QCD. This program was finally
accomplished by two independent groups. One in Munich around A.~Buras
and one in Rome around G.~Martinelli. 

\section{$K\to\pi\pi$ and the $\Delta I=1/2$ rule}
\label{deltaIhalf}

\begin{figure}[t]
\begin{minipage}{0.66\textwidth}
\centerline{\includegraphics[width=0.99\textwidth]{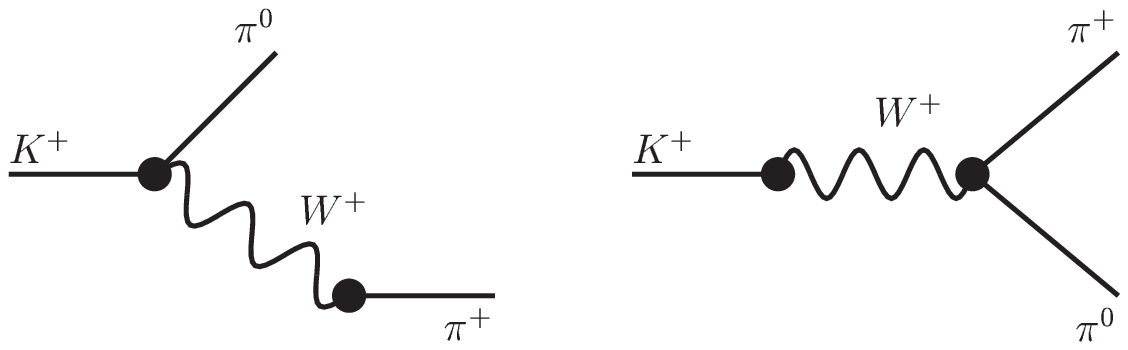}}
\centerline{(a)}
\end{minipage}
\begin{minipage}{0.33\textwidth}
\centerline{\includegraphics[width=0.99\textwidth]{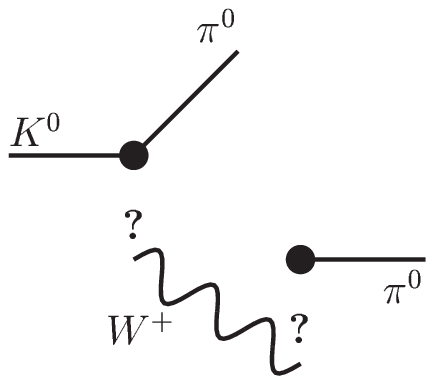}}
\centerline{(b)}
\end{minipage}
\caption{(a) The two naive $W^+$-exchange diagrams for
$K^+\longrightarrow \pi^+\pi^0$.
(b) No simple $W^+$-exchange diagram is possible for
$K^0\longrightarrow \pi^0\pi^0$.}
\label{figkpipiplus}
\end{figure}

The underlying qualitative difference we want to understand is the
$\Delta I=1/2$ rule.
We can try to calculate $K\to\pi\pi$
decays by simple $W^+$ exchange.
For $K^+\to\pi^+\pi^0$ we can draw the two Feynman diagrams
of Fig.~\ref{figkpipiplus}(a).
The $W^+$-hadron couplings are known from semi-leptonic
decays. This approximation agrees with the
measured decay within a factor of two.

A much worse result appears when we try the same for
$K^0\to\pi^0\pi^0$. As shown in Fig. \ref{figkpipiplus}(b) there is no
possibility to draw diagrams similar to those in Fig. \ref{figkpipiplus}(a).
The needed vertices always violate charge-conservation.
So we expect that the neutral decay should be small compared
with the ones with charged pions. Well, if we look at the experimental
results, we see
\ba
\Gamma(K^0\longrightarrow\pi^0\pi^0)=&
\dsp\frac{1}{2}\Gamma(K_S\longrightarrow\pi^0\pi^0)&=2.3\cdot10^{-12}\mbox{ MeV}
\nonumber\\
\Gamma(K^+\longrightarrow\pi^+\pi^0)=&\dsp
1.1\cdot10^{-14}\mbox{ MeV}&
\ea
So the expected zero one is by far the largest !!!

The same conundrum can be expressed in terms of the
 isospin amplitudes:\,\footnote{The  sign
convention is the one used in the work by J.~Prades and myself.}
\ba
A[K^0 \to \pi^0 \pi^0]  &\equiv& \sqrt{{1}/{3}} { A_0}
-\sqrt{{2}/{3}} \, { A_2} \nonumber\\ 
A[K^0 \to \pi^+ \pi^-]  &\equiv& \sqrt{{1}/{3}} { A_0}
+\sqrt{{1}/{6}} \, { A_2} \nonumber\\ 
A[K^+ \to \pi^+ \pi^0]  &\equiv& ({\sqrt{3}}/{2}) { A_2}\,.
\ea
The above quoted experimental results can now be rewritten as
\be 
\label{dIhalf1}
\left|{A_0}/{A_2}\right|_{\mbox{exp}} = 22
\ee
while the naive $W^+$-exchange discussed would give
\be
\label{dInaive}
\left|{A_0}/{A_2}\right|_{\mbox{naive}} = \sqrt{2}\,.
\ee
This discrepancy is known as the problem of the $\Delta I=1/2$ rule.

Some enhancement comes from final state $\pi\pi$-rescattering.
Removing these and higher order effects in the light quark masses
one obtains~\cite{KMW,BDP}
\be
\label{dIhalf2}
\dsp\left|{A_0}/{A_2}\right|_{\chi} = 17.8\,.
\ee
This changes the discrepancy somewhat but is still different by an order
of magnitude from the naive result (\ref{dInaive}). The difference
will have to be explained by pure strong interaction effects
and it is a {\em qualitative} change, not just a quantitative one.

We also use amplitudes without the final state interaction
phase:
\be
A_I=-ia_I e^{i\delta_I}
\ee
for $I=0,2$. $\delta_I$ is the angular momentum zero, isospin I scattering
phase at the Kaon mass.

\section{$K\to\pi\pi$, $\varepsilon$, \epspeps}
\label{epspeps}

The $K^0$, $\kob$\ states have $\bar sd$,  $\bar ds$
quark content. $CP$ acts on these states as
\be
CP |K^0\rangle = -|\kob\rangle\,.
\ee
{}We can construct eigenstates with a definite $CP$ transformation:
\be
K^0_{1(2)} = \frac{1}{\sqrt{2}}
\left(K^0 -(+) \kob\right),\quad\quad
 CP|K_{1(2)} = +(-)|K_{1(2)}\,.
\ee
The main decay mode of $K^0$-like states is $\pi\pi$. A two pion state
with charge zero in spin zero is always CP even.
Therefore the decay $K_1\to\pi\pi$ is possible
but $K_2\to\pi\pi$ is {\em impossible}; $K_2\to\pi\pi\pi$ is possible.
Phase-space for the $\pi\pi$ decay is much larger than for the
three-pion final state. Therefore if we start out with a pure $K^0$ or $\kob$\
state, the
$K_2$ component in its wave-function lives much longer than the $K_1$
component such that after a long time only the $K_2$ component
survives.

In the early sixties, as you see it pays off to do precise experiments,
one actually measured \cite{CCFT}
\be
\frac{\Gamma(K_L\to\pi^+\pi^-)}{\Gamma(K_L\to\mbox{all})}=
(2\pm0.4)\cdot10^{-3}\ne 0\,,
\ee
showing that {\em $CP$ is violated}\,.
This leaves us with the questions:
\begin{itemize}
\item[{\bf ???}] Does $K_1$ turn in to $K_2$
(mixing or indirect $CP$ violation)?
\item[{\bf ???}] Does $K_2$ decay directly into $\pi\pi$ 
(direct $CP$ violation)?
\end{itemize}
In fact, the 
answer to both is {\em YES}\, and is major qualitative test
of the standard model Higgs-fermion sector and the $CKM$-picture
of $CP$-violation.

The presence of $CP$-violation means that $K_1$ and $K_2$ are not the mass
eigenstates, these are
\be
K_{S(L)} = \frac{1}{\sqrt{1+|\tilde\varepsilon|^2}}
\left(K_{1(2)}+\tilde\varepsilon K_{2(1)}\right)\,.
\ee
They are not orthogonal since the Hamiltonian is not hermitian.

We define the observables
\ba
\varepsilon &\equiv& \frac{A(K_L\to(\pi\pi)_{I=0})}{A(K_S\to(\pi\pi)_{I=0})}
\nonumber\\
\varepsilon^\prime &=& \frac{1}{\sqrt{2}}\left(
\frac{A(K_L\to(\pi\pi)_{I=2})}{A(K_S\to(\pi\pi)_{I=0})}-
\varepsilon
\frac{A(K_S\to(\pi\pi)_{I=2})}{A(K_S\to(\pi\pi)_{I=0})}\right)\,.
\ea
The latter has been specifically constructed to remove the 
$K^0$-$\kob$\ transition.
$|\varepsilon|$ is a directly measurable as ratios of decay rates.

We now make a series of experimentally valid approximations,
\be
 |\im a_0|,|\im a_2| << |\re a_2| << |\re a_0|,\quad
 |\varepsilon|,|\tilde \varepsilon| << 1,
 \quad |\varepsilon^\prime| << | \varepsilon|\,,
\ee
to obtain the usually quoted expression
\be
\label{epsp}
\varepsilon^\prime = \frac{i}{\sqrt{2}}e^{i(\delta_2-\delta_0)}
\frac{\re a_2}{\re a_0}\left(\frac{\im a_2}{\re a_2}-\frac{\im a_0}{\re a_0}
\right)\,.
\ee

Experimentally,\cite{PDG}
\be
|\varepsilon| = (2.271\pm0.017)\cdot 10^{-3}\,.
\ee

\begin{figure}[t]
\begin{minipage}{0.45\textwidth}
\centerline{\includegraphics[width=0.99\textwidth]{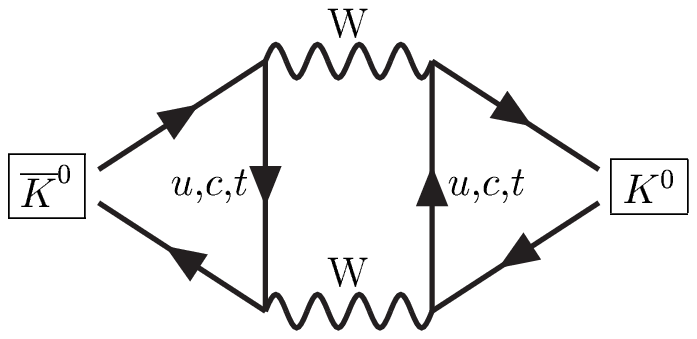}}
\centerline{(a)}
\end{minipage}
\begin{minipage}{0.54\textwidth}
\centerline{\includegraphics[width=0.95\textwidth]{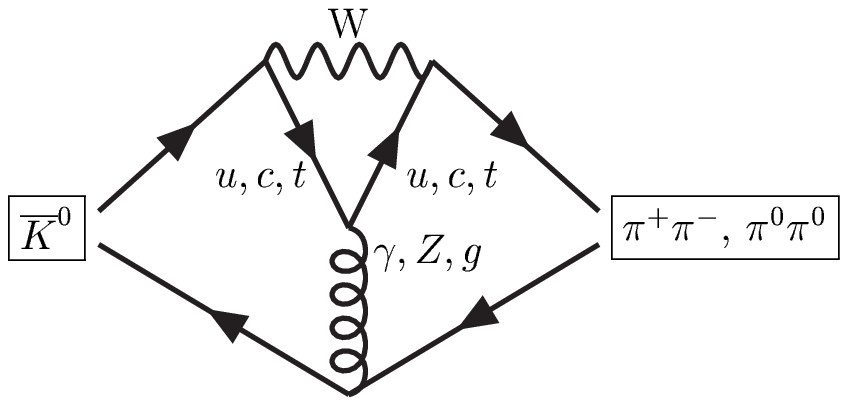}}
\centerline{(b)}
\end{minipage}
\caption{\label{figbox}
(a) The box diagram contribution to $K^0\kob$\ mixing.
Crossed versions and diagrams with extra gluons etc. are not shown.
(b) The Penguin diagram contribution
to $K\to\pi\pi$.
Extra gluons and crossed versions etc. are not shown.}
\end{figure}

The set of diagrams, depicted schematically in Fig.~\ref{figbox}(a),
responsible for $K^0\kob$\ mixing are known as box diagrams.
It is the presence of the virtual intermediate quark lines
of up, charm and top quarks that produces the $CP$-violation. 

The experimental situation on $\epspeps$ was unclear for a long time.
Two large experiments, NA31 at CERN and E731 at FNAL,
obtained conflicting results
in the mid 1980's. Both groups have since gone on and build improved versions
of their detectors, NA48 at CERN and KTeV at FNAL.
$\epspeps$ is
measured via the double ratio
\be
\re\left(\frac{\varepsilon^\prime}{\varepsilon}\right)
= \frac{1}{6}\left\{
1-\frac{\dsp \Gamma(K_L\to\pi^+\pi^-)/\Gamma(K_S\to\pi^+\pi^-)}
{\dsp \Gamma(K_L\to\pi^0\pi^0)/\Gamma(K_S\to\pi^0\pi^0)}
\right\}\,.
\ee
The two main experiments follow a somewhat different strategy in
measuring this double ratio, mainly in the way the relative normalisation
of $K_L$ and $K_S$ components is treated.
After some initial disagreement with the first results, KTeV has
reanalysed their systematic errors and the
situation for $\epspeps$ is now quite clear.
We show the recent results in Table \ref{tabepspeps}.
The data are taken from Ref.~\citebk{epspeps} and the recent
reviews in the Lepton-Photon conference.\cite{Kessler,NA4801} 
\begin{table}[t]
\tbl{Recent results on $\varepsilon^\prime/\varepsilon$. The years
refer to the data sets.}
{\footnotesize
{\begin{tabular}{|cc|}
\hline
NA31 & $(23.0\pm6.5)\times 10^{-4}$\\
E731 & $(7.4\pm5.9)\times 10^{-4}$\\
\hline
KTeV 96 & $(23.2\pm4.4)\times10^{-4}$\\
KTeV 97 & $(19.8\pm2.9)\times10^{-4}$\\
NA48 97 & $(18.5\pm7.3)\times10^{-4}$\\
NA48 98+99 & $(15.0\pm2.7)\times10^{-4}$\\
\hline
ALL & $(17.2\pm1.8)\times10^{-4}$\\
\hline
\end{tabular}}}
\label{tabepspeps}
\end{table}

The Penguin diagram shown in Fig. \ref{figbox}(b) contributes to the direct
$CP$-violation as given by $\varepsilon^\prime$.
Again, $W$-couplings to all three generations show up so $CP$-violation
is possible in $K\to\pi\pi$. This is a qualitative prediction of the
standard model and borne out by experiment.
The main problem is now to embed these diagrams and the simple $W$-exchange
in the full strong interaction. The $\Delta I=1/2$ rule shows that there
will have to be large corrections to the naive picture.

\section{From Quarks to Mesons: a Chain of Effective Field Theories}
\label{EFT}

The full calculation in the presence of the strong interaction is quite
difficult. Even at short distances, due to the presence of logarithms
of large ratios of scales, a simple one-loop calculation gives
very large effects. These need to be resummed which fortunately can be done
using renormalisation group methods. 

The three steps of the
full calculation are depicted in 
Fig. \ref{figsteps}.
\begin{figure}
\begin{tabular}{ccc}
ENERGY SCALE & FIELDS & Effective Theory\\
\hline
\\[2mm]
$M_W$ & \framebox{\parbox{5cm}{\begin{center}
$W,Z,\gamma,g$;\\$\tau,\mu,e,\nu_\ell$;\\
$t,b,c,s,u,d$\end{center}
}} & Standard Model\\[2mm]
 & {\em $\Downarrow$ using OPE} & \\[2mm]
$\lesssim m_c$ & \framebox{\parbox{5cm}{\begin{center}
$\gamma,g$; $\mu,e,\nu_\ell$;\\ $s,d,u$\end{center}
}} & \parbox{2cm}{QCD,QED,\\${\mathcal H}_{\mbox{eff}}^{|\Delta S|=1,2}$}\\[2mm]
  & {\em$\Downarrow$ ???} & \\[2mm]
$M_K$  & \framebox{\parbox{5cm}{\begin{center}$\gamma$; $\mu,e,\nu_\ell$;\\
$\pi$, $K$, $\eta$\end{center}}} & CHPT\\ \\\hline
\end{tabular}
\caption{A schematic exposition of the various steps in the calculation
of nonleptonic matrix-elements.}
\label{figsteps}
\end{figure}
First we integrated out the heaviest particles step by step using Operator
Product Expansion methods. The steps OPE we describe in the next subsections
while step {\em ???} we will split up in more subparts later.

\subsection{Step I: from SM to OPE}

\begin{figure}[t]
\begin{minipage}{0.495\textwidth}
\centerline{\includegraphics[width=0.999\textwidth]{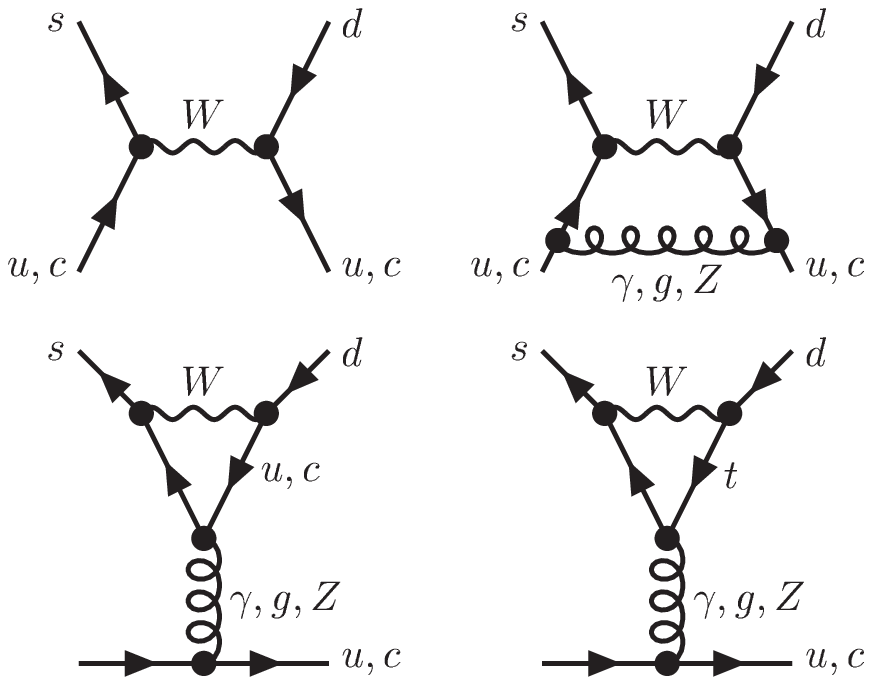}}
\centerline{(a)}
\end{minipage}
\begin{minipage}{0.495\textwidth}
\centerline{\includegraphics[width=0.999\textwidth]{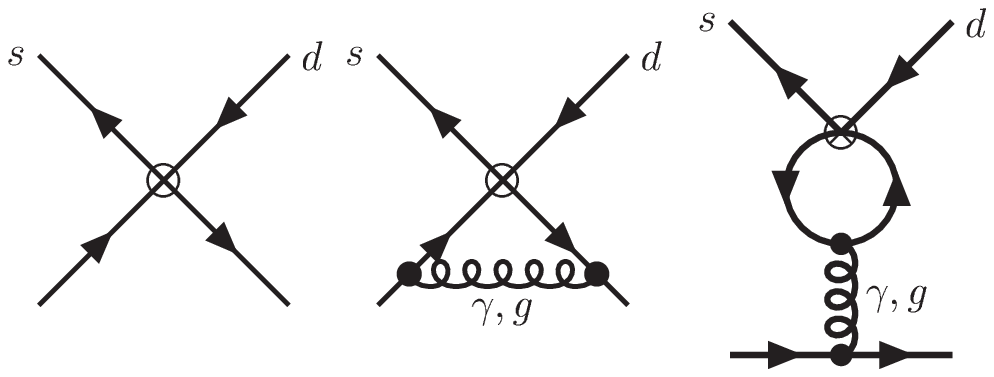}}
\centerline{(b)}
\end{minipage}
\caption{(a) The standard model diagrams to be calculated at a high scale.
(b) The diagrams needed for the
matrix-elements calculated at a scale $\mu_H\approx m_W$
using the effective Hamiltonian.}
\label{figSM}
\end{figure}

The first step concerns the
standard model diagrams of Fig. \ref{figSM}(a).
We replace their effect
with a contribution of an effective Hamiltonian given by
\be
\label{defHeff}
 {\mathcal H}_{\mbox{eff}} = \sum_i C_i(\mu) Q_i(\mu)
= \frac{G_F}{\sqrt{2}} V_{ud}V_{us}^*
\sum_i\left(z_i-y_i\frac{V_{td}V_{ts}^*}{V_{ud}V_{us}^*}\right)Q_i\,.
\ee
In the last part we have real coefficients $z_i$ and $y_i$ and
the CKM-matrix-elements occurring are shown explicitly.
The four-quark operators $Q_i$ can be found in e.g. Ref.~\citebk{BPscheme}.

We calculate now matrix-elements between quarks and gluons in the
standard model using the diagrams of Fig. \ref{figSM}(a)
and equate those to the same matrix-elements calculated using the
effective Hamiltonian of Eq. (\ref{defHeff}) and the diagrams
of Fig. \ref{figSM}(b). This determines the value of the $z_i$ and $y_i$.
The top quark and the $W$ and $Z$ bosons are integrated out all at the
same time. There should be no large logarithms present due to that.
The scale $\mu = \mu_H$ in the diagrams of Fig.~\ref{figSM}(b)
of the OPE expansion diagrams
should be chosen of the order of the $W$ mass.
The scale $\mu_W$ in the Standard Model diagrams of Fig.~\ref{figSM}(a)
should be chosen of the same order.

{\underline{Notes}:}\\
$\bullet$ In the Penguin diagrams
$CP$-violation shows up since all 3 generations
are present.\\
$\bullet$ The equivalence is done by calculating matrix-elements between
{\em Quarks and Gluons}\\
$\bullet$ The SM part is $\mu_W$-independent to $\alpha_S^2(\mu_W)$.\\
$\bullet$ OPE part: The $\mu_H$ dependence of
$C_i(\mu_H)$ cancels the $\mu_H$ dependence of the diagrams
to order $\alpha_S^2(\mu_H)$.

This procedure gives at $\mu_W = \mu_H=M_W$
in the NDR-scheme~\footnote{The precise definition
of the four-quark operators $Q_i$ comes in here as well. See the lectures by
Buras \cite{Buras1} for a more extensive description of that.}
the numerical values given in Table \ref{tabzimw}.
\begin{table}[t]
\tbl{The Wilson coefficients and their main source at
the scale $\mu_H=m_W$ in the NDR-scheme.}
{\footnotesize
\begin{tabular}{|ccc|ccc|}
\hline
$z_1$   & 0.053     &  { $g,\gamma$-box}               &
$y_6$	&$-$0.0019  & {$g$-Penguin}\\
$z_2$	& 0.981	    &{ $W^+$-exchange  $g,\gamma$-box} &
$y_7$	& 0.0009    & {$\gamma,Z$-Penguin}\\
$y_3$	& 0.0014    & {$g,Z$-Penguin  $WW$-box}        &
$y_8$	& 0.	    & \\
$y_4$	&$-$0.0019  & {$g$-Penguin}                    &
$y_9$	& $-$0.0074 & {$\gamma,Z$-Penguin  $WW$-box}\\
$y_5$	& 0.0006    & {$g$-Penguin}                    &
$y_{10}$& 0.        & \\
\hline
\end{tabular}
}
\label{tabzimw}
\end{table}
In the same table I have
given the main source of these numbers. Pure tree-level $W$-exchange
would have only given $z_2=1$ and all others zero.
Note that the coefficients from $\gamma,Z$ exchange are similar to the gluon
exchange ones since $\alpha_S$ at this scale is not very big.

\subsection{Step II}
\label{shortdistance}

Now comes the main advantage of the OPE formalism. Using the
renormalisation group equations we can  calculate
the change with $\mu$ of the $C_i$, thus resumming the
$\log\left(m_W^2/\mu^2\right)$ effects.
The renormalisation group equations (RGEs) for the
strong coupling and the Wilson coefficients are
\be
\label{RGE}
\mu\frac{d}{d\mu}g_S(\mu) = \beta(g_S(\mu)),\quad
\mu\frac{d}{d\mu}C_i(\mu) = \gamma_{ji}(g_S(\mu),\alpha) C_j(\mu)\,.
\ee
$\beta$ is the QCD beta function for the running coupling.
The coefficients
$\gamma_{ij}$ are the elements
of the anomalous dimension matrix $\hat\gamma$. They can be derived from
the infinite parts of loop diagrams and this has been done to
one~\cite{one-loop} and two loops.\cite{two-loop}
The series in $\alpha$ and $\alpha_S$ is known to
\be
\hat\gamma = \hat\gamma^0_S \frac{\alpha_S}{4\pi} 
            + \hat\gamma^1_S \left(\frac{\alpha_S}{4\pi}\right)^2
            + \hat\gamma_e \frac{\alpha}{4\pi}
            + \hat\gamma_{se}\frac{\alpha_S}{4\pi} \frac{\alpha}{4\pi}
            +\cdots 
\ee
Many subtleties are involved in this calculation.\cite{Buras1,two-loop}
They all are related to the fact that everything at higher loop orders need
to be specified correctly, and many things which are equal at tree
level are no longer so in $d\ne4$ and at higher loops, see the
lectures~\citebk{Buras1} or the review~\citebk{twoloopreview}.
The numbers below are obtained by numerically
integrating Eq. (\ref{RGE}).\cite{BPdIhalf,BPeps}

We perform the following steps to get down to a scale $\mu_{OPE}$
 around 1~GeV. Starting from the $z_i$ and $y_i$
at the scale {\boldmath$\mu_H$}:\\
(1)
solve Eqs. (\ref{RGE}); run from {\boldmath${\mu_H}$}
to {\boldmath${\mu_b}$}.\\
(2)
At {\boldmath$\mu_b$}$\approx m_b$ remove $b$-quark and match to the
theory without $b$ by calculating matrix-elements of the
effective Hamiltonian in the five and in the four-quark picture
and putting them equal.\\
(3)
Run from {\boldmath$\mu_b$} to {\boldmath$\mu_c$}$\approx m_c$.\\
(4)
At {\boldmath$\mu_c$} remove the $c$-quark and match to
the theory without $c$.\\
(5)
Run from {\boldmath$\mu_c$} to {\boldmath$\mu_{\mbox{OPE}}$}.\\
Then {\em all}\, large logarithms including $m_W$, $m_Z$, $m_t$,
$m_b$ and $m_c$, are summed.

With the inputs
$m_t(m_t)=166~GeV$, $\alpha=1/137.0$, $\alpha_S(m_Z) = 0.1186$
which led to the initial conditions shown in Table~\ref{tabzimw}, we can
perform the above procedure down to $\mu_{OPE}$.
Results for 900~MeV are shown in columns two and three of Table \ref{tabzimu}.
\begin{table}[t]
\tbl{The Wilson coefficients
$z_i$ and $y_i$ at a scale $\mu_{\mbox{OPE}}=$ 900~MeV
in the NDR scheme and in the $X$-boson scheme at $\mu_X =$ 900~MeV.}
{\footnotesize
\begin{tabular}{|c|cc|cc|}
\hline
i        &\hspace{6mm} $z_i$\hspace{6mm} &\hspace{6mm} $y_i$\hspace{6mm} &\hspace{6mm} $z_i$\hspace{6mm} &\hspace{6mm} $y_i$\hspace{6mm} \\
\hspace{1cm}         &\multicolumn{2}{c}{$\mu_{OPE} = 0.9$ GeV}\vline
&\multicolumn{2}{c}{$\mu_X =0.9$ GeV}\vline\\
\hline
$z_{1}$ &$-$0.490           & 0.                & $-$0.788        & 0.\\
$z_{2}$	&1.266 		    & 0.                & 1.457           & 0. \\
$z_{3}$	&0.0092		    &  0.0287           & 0.0086          & 0.0399\\
$z_{4}$	&$-$0.0265	    & $-$0.0532         & $-$0.0101       & $-$0.0572\\
$z_{5}$	& 0.0065	    & 0.0018            & 0.0029          & 0.0112\\
$z_{6}$	&$-$0.0270	    & $-$0.0995         &  $-$0.0149      & $-$0.1223\\
$z_{7}$	& 2.6$~10^{-5}$	    &$-$0.9$~10^{-5}$   & 0.0002          &$-$0.00016\\
$z_{8}$	& 5.3$~10^{-5}$	    & 0.0013            &  6.8$~10^{-5}$  & 0.0018\\
$z_{9}$	& 5.3$~10^{-5}$	    & $-$0.0105         &  0.0003         & $-$0.0121\\
$z_{10}$&$-$3.6$~10^{-5}$   &0.0041             & $-$8.7$~10^{-5}$& 0.0065\\
\hline
\end{tabular}
}
\label{tabzimu}
\end{table}
$z_1$ and $z_2$ have changed much from $0$ and $1$.
This is the short-distance contribution to the $\Delta I=1/2$ rule.
We also see a large enhancement of $y_6$ and $y_8$, which will
lead to our value of $\varepsilon^\prime$.

\subsection{Step III: Matrix-elements}
\label{longdistance1}

Now
remember that the $C_i$ depend on $\mu_{OPE}$ (scale dependence)
and on the definition of the $Q_i$ (scheme dependence)
and the numerical change in the coefficients
due to the various choices for the $Q_i$ possible is not negligible.
It is therefore important both from the phenomenological and fundamental
point of view that this dependence is correctly accounted for in
the evaluation of the matrix-elements.
We can solve this in various ways.\\
$\bullet$ {\bf Stay in QCD} $\Rightarrow$
Lattice calculations.\cite{Sachrajda}\\
$\bullet$ {\bf ITEP Sum Rules} or QCD sum rules.~\cite{SVZ}\\
$\bullet$ {\bf Give up} $\Rightarrow$ Naive factorisation.\\
$\bullet$ {\bf Improved factorisation}\\
$\bullet$ {\bf  $X$-boson method} (or fictitious Higgs method)\\
$\bullet$ {\bf  Large $N_c$} (in combination with something like
the $X$-boson method.) Here the difference is mainly in the
treatment of the low-energy hadronic physics. Three main approaches
exist of increasing sophistication.\footnote{Which of course means that
calculations exist only for simpler matrix-elements for the more sophisticated
approaches.}
\begin{flushright}
\begin{minipage}{0.97\textwidth}
$\circledast$ CHPT: As originally proposed by Bardeen-Buras-G\'erard~\cite{BBG}
and now pursued mainly by Hambye and collaborators.\cite{Hambye}\\
$\circledast$ ENJL (or extended Nambu-Jona-Lasinio model~\cite{ENJL}):
As mainly done by
myself and J.~Prades.\cite{BPBK,BPscheme,kptokpp,BPdIhalf,BPeps}\\
$\circledast$ LMD or lowest meson dominance approach.\cite{LMD} 
These papers stay
with dimensional regularisation throughout. The $X$-boson corrections
discussed below, show up here as part of the QCD corrections.
\end{minipage}
\end{flushright}
$\bullet$ {\bf Dispersive methods} Some matrix-elements can in principle
be deduced
from experimental spectral functions.

Notice that there other approaches as well, e.g.~the chiral quark
model.\cite{PichdeRafael} 
These have no underlying arguments why the $\mu$-dependence
should cancel, but the importance of several effects
was first discussed in this context. I will also not treat the
calculations done using bag models and potential models which
similarly do not address the $\mu$-dependence issue.

\section{The $X$-boson Method and Results using ENJL for the Long Distance}
\label{Xboson}

We want to have a consistent calculational scheme that takes the scale
and scheme dependence into account correctly. Let us therefore have a closer
look at how we calculate the matrix-elements using naive factorisation.
We start from the four-quark operator:\\
$\Rightarrow$ See it as a product of currents or densities.\\
$\Rightarrow$ Evaluate current matrix-elements in low energy
theory or model or from experiment.\\
$\Rightarrow$ Neglect extra momentum transfer between the current matrix
elements.\\
The main lesson here is that {\em currents and densities are easier
to deal with.} We also need to go beyond the approximation in the last step.
 To obtain well defined currents, we replace the four-quark operators
by exchanges of fictitious massive $X$-bosons
coupling to two-quark currents or densities.
\be
{\mathcal H} = \sum_i C_i(\mu_{OPE}) Q_i
\quad
\Longrightarrow
\quad
\sum_i g_i X_i J_i\,.
\ee
$\bullet$ This is a well defined scheme of nonlocal operators.\\
$\bullet$ The matching to obtain the coupling constants
$g_i$ from the $C_i$ is done with
matrix-elements of {\em quarks and gluons}.

A simple example is the one needed for the $B_K$ parameter. The
four-quark operator is replaced by the exchange of one $X$-boson $X_B$:
\be
C(\mu) \bar d\gamma_\mu(1-\gamma_5)s\,\bar d\gamma_\mu(1-\gamma_5)s
\quad
\Longrightarrow
\quad
g_B X_B^\mu \bar d\gamma_\mu(1-\gamma_5)s\,.
\ee
Taking a matrix-element between quarks at
next-to-leading order in $\alpha_S$
gives
\be
\label{Xmatching}
C(\mu_{OPE}) \left(1+ r\alpha_S(\mu_{OPE})\right) =
(g_B^2/M_{X_B}^2) \left(1+r'\alpha_S+r''\log\left(M_{X_B}^2/\mu^2\right)\right)
\,.
\ee
The coefficients $r$ and $r'$ take care of the scheme dependence.
The l.h.s. is scale independent to the required order in $\alpha_S$.
The effect of these coefficients surprisingly always went in the
direction to improve agreement with experiment~\cite{BPscheme,BPeps}
as can be seen from columns 4 and 5 in Table~\ref{tabzimu}.

The final step is the matrix-element of $X_B$-boson exchange.
For this we split the integral over the  $X_B$ momentum $q_X$ in two parts
\be
\int dq_X^2 \Longrightarrow
\int_0^{\mu^2} dq_X^2 +\int_{\mu^2}^\infty dq_X^2\,.
\ee
\begin{figure}[t]
\centerline{\includegraphics[width=0.8\textwidth]{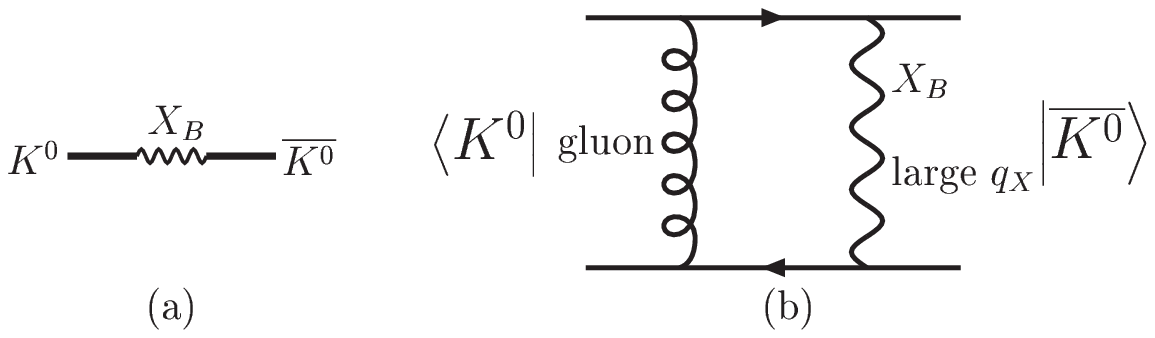}}
\caption{(a) The leading in $1/N_c$ contribution from $X_B$ exchange.
(b) The large momentum part of the $X_B$ exchange matrix-element.}
\label{figX}
\end{figure}
The leading in $N_c$ contribution is depicted in Fig.~\ref{figX}(a)
and corresponds to the large $N_c$ factorisation.
The large momentum regime is evaluated by the diagram in Fig.~\ref{figX}(b),
since the large momentum must flow back through quarks and gluons.
Hadronic exchanges are power suppressed because of the form factors involved.
The $\alpha_S$ present already suppresses by $N_c$ so the matrix-element
of this part can be evaluated using factorisation.
This part cancels the $r''\log(M_{X_B}^2/\mu^2)$ present
in (\ref{Xmatching}). The final part with small $q_X$ momentum in the
integral then needs to be evaluated nonperturbatively.
Here one can use Chiral Perturbation Theory, the ENJL model
or meson exchange approximations with various short-distance constraints.
 
\begin{figure}[t]
\centerline{\includegraphics[angle=270,width=0.495\textwidth]{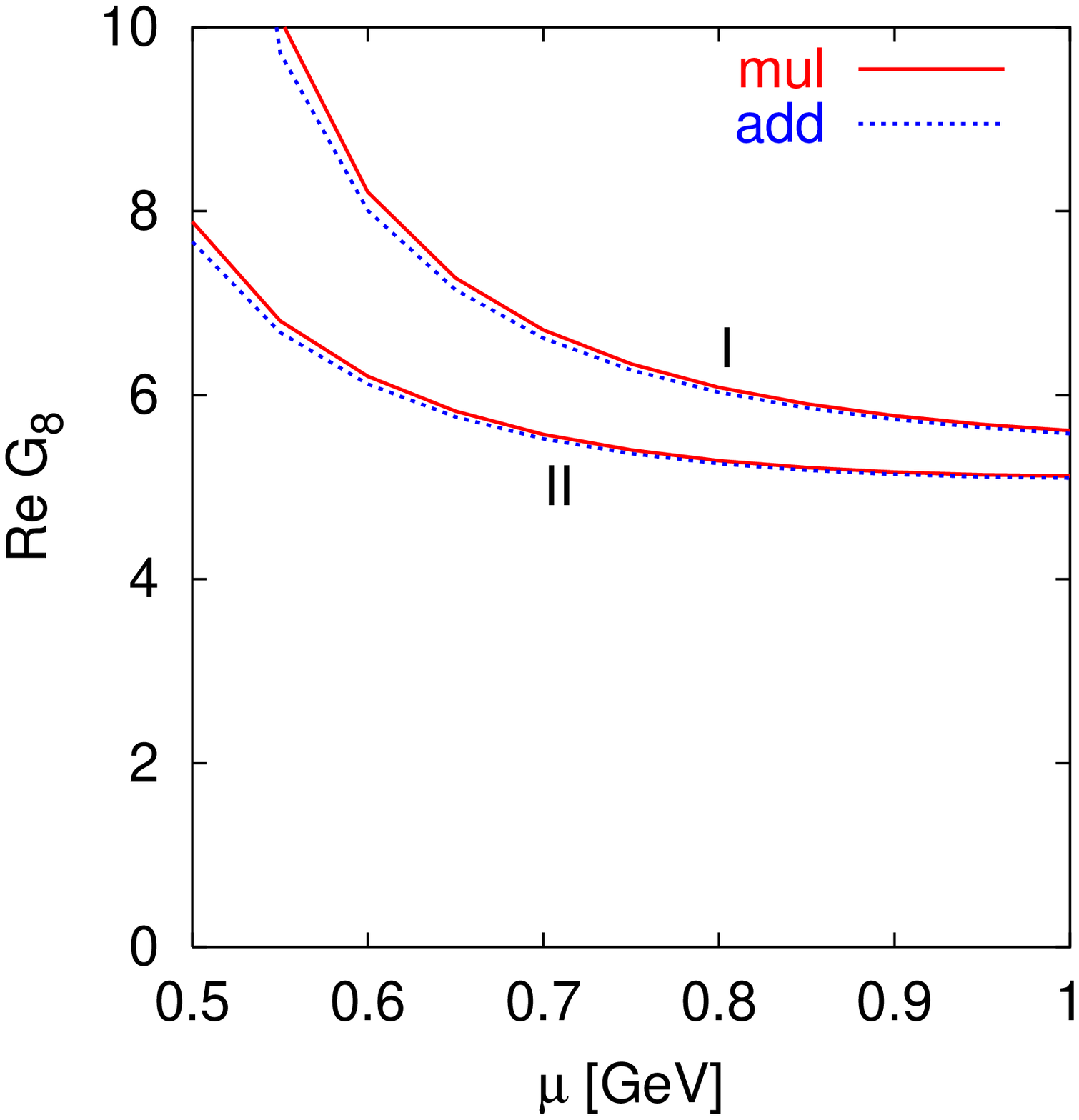}
\includegraphics[angle=270,width=0.495\textwidth]{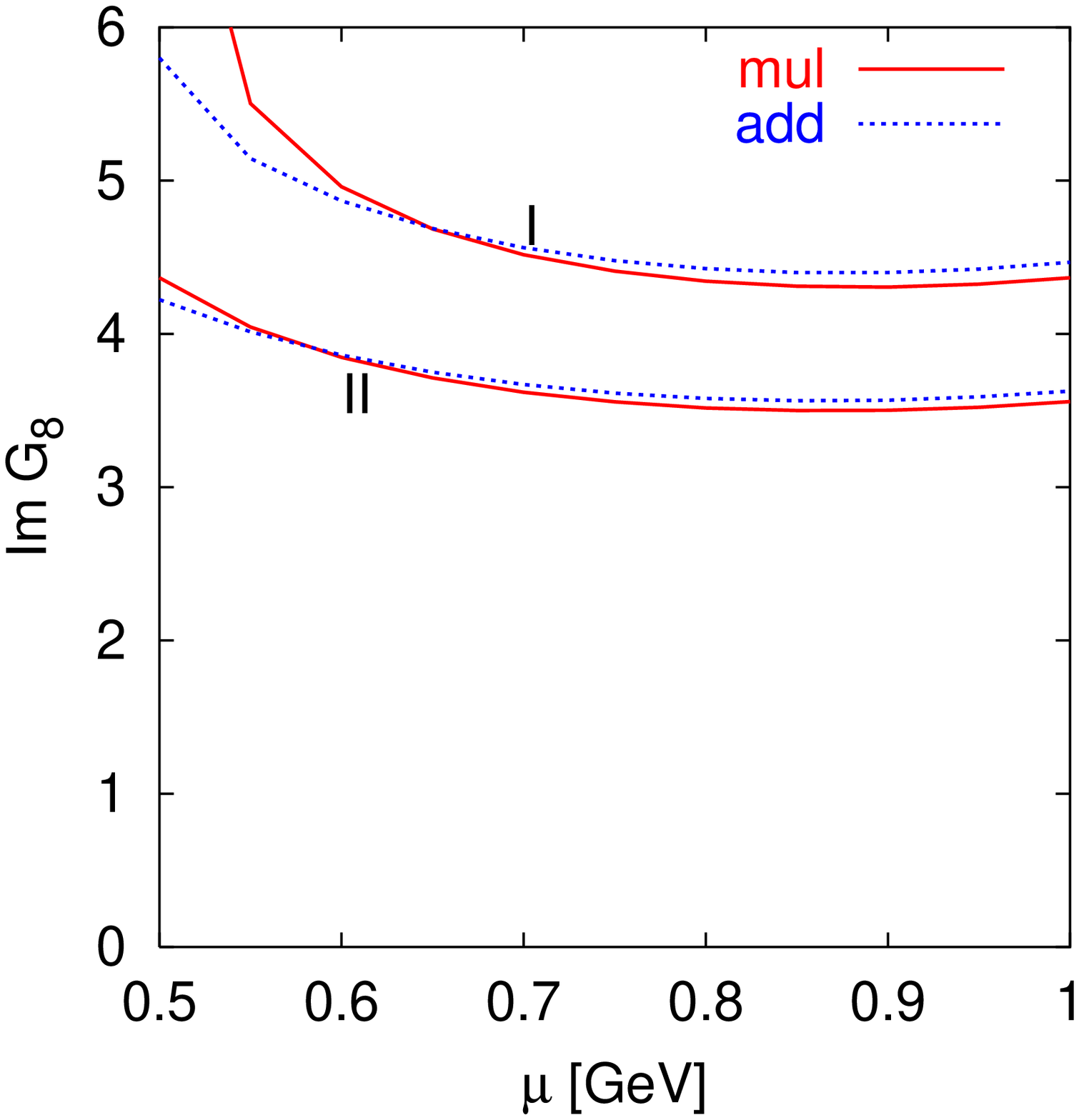}
}
\caption{Left: the results for the real part of the octet $\re G_8$
as a function of $\mu$. Right: the same for the imaginary part.}
\label{figG8}
\end{figure}

Let me now show some results from Refs~\citebk{BPscheme,BPeps}.
The chiral limit coupling $G_8$ responsible for the octet contribution,
it is 1 in the naive approximation and about 6 when fitted to
experiment,\cite{KMW,BDP} is shown in Fig.~\ref{figG8}.
As can be seen, the matching between the short and long distance is reasonable
both for the real and imaginary parts.
The value of $\re G_8$ is dominated by the matrix-element of $Q_1$ and $Q_2$
but about 30-60\% comes from the long distance Penguin part of $Q_2$.
The result for the $\im G_8$ corresponds to a value of the $Q_6$ 
matrix-element much larger than usually assumed. We obtained $B_6\approx2$-2.5
while it is usually {\em assumed} to be less than 1.5.

Putting our results in (\ref{epsp}) we obtain a chiral limit value
for $(\epspeps)_\chi$ of about $6\cdot 10^{-3}$.

We now add the main isospin breaking component $\Omega$~\cite{EPetal}
and the effect of final state interaction (FSI).\cite{PP}
The latter in our case has mainly effect on the forefactor $\re a_2/\re a_0$
in (\ref{epsp}) since the ratios of imaginary parts have been evaluated
to the same order in $p^2$ in CHPT and thus receive no FSI corrections.
The final result is~\cite{BPeps} $\epspeps\approx 1.5\cdot 10^{-3}$
with an error $\gtrapprox$ 50\%.

\section{Dispersive Estimates for $\langle Q_7\rangle$ and
$\langle Q_8\rangle$}
\label{dispersive}

Some of the matrix-elements we want can be extracted from
experimental information in a different way. The canonical
example is the mass difference between the charged and the neutral pion
in the chiral limit which can be extracted from a dispersive integral
over the difference of the vector and axial vector spectral
functions.\cite{Dasetal}

This idea has been pursued in the context of weak decay in a series of
papers by Donoghue, Golowich and collaborators.\cite{DG1}
The matrix-element of $Q_7$ could be extracted directly
from these data. To get at the matrix-element of $Q_8$ is somewhat more
difficult. Ref. \citebk{DG1} extracted it first
by requiring $\mu$-independence, this corresponds
to extracting  the matrix
element of $Q_8$ from the spectral functions
via the coefficient of the dimension 6 term in the operator
product expansion of the underlying Green's function.
The most recent papers using this method are
Refs.~\citebk{KnechtQ7Q8,NAR01,DG2} and \citebk{BPQ7Q8}.
In the last two papers also some QCD corrections were included which had a
substantial impact on the numerical results.

The results are given in Table.~\ref{tableQ7Q8}. The operator
$O_6^{(1)}$ is related by a chiral transformation to $Q_7$
and $O_6^{(2)}$ to $Q_8$.
The numbers are valid in the chiral limit.
\begin{table}[t]
\tbl{The values of the VEVs in the NDR scheme at $\mu_R=2$ GeV.
The most recent dispersive results are line 3 to 5.
The other results are shown for comparison. Errors are those quoted in the
papers.
Adapted from Ref.~[71].
}
{\footnotesize
\begin{tabular}{|c|c|c|}
\hline
Reference&$\langle 0| O_6^{(1)}|0 \rangle^{NDR}_\chi$&
$\langle 0| O_6^{(2)}|0 \rangle^{NDR}_\chi$ 
 \\ 
\hline
$B_7=B_8=1$                      & $-(5.4\pm2.2)\cdot10^{-5}$ GeV$^6$ & $(1.0 \pm 0.4)\cdot10^{-3}$ GeV$^6$   \\
\hline				   
Bijnens et al. \citebkcap{BPQ7Q8}     &$-(4.0\pm0.5)\cdot10^{-5}$ GeV$^6$& $(1.2  \pm 0.5)\cdot10^{-3}$ GeV$^6$  \\  
Knecht et al. \citebkcap{KnechtQ7Q8}  &$-(1.9\pm0.6)\cdot10^{-5}$ GeV$^6$& $(2.3 \pm 0.7)\cdot10^{-3}$ GeV$^6$   \\   
Cirigliano et al. \citebkcap{DG2}     &$-(2.7\pm1.7)\cdot10^{-5}$ GeV$^6$& $(2.2 \pm 0.7)\cdot10^{-3}$ GeV$^6$   \\  
\hline				   
Donoghue et al.\citebkcap{DG1}        &$-(4.3\pm0.9)\cdot10^{-5}$ GeV$^6$& $(1.5 \pm 0.4)\cdot10^{-3}$ GeV$^6$      \\  
Narison        \citebkcap{NAR01}      &$-(3.5\pm1.0)\cdot10^{-5}$ GeV$^6$& $(1.5 \pm 0.3)\cdot10^{-3}$ GeV$^6$      \\  
lattice   \citebkcap{Q7Q8lattice}     &$-(2.6\pm0.7)\cdot10^{-5}$ GeV$^6$& $(0.74 \pm 0.15)\cdot10^{-3}$ GeV$^6$    \\
ENJL  \citebkcap{BPeps}               &$-(4.3\pm0.5)\cdot10^{-5}$ GeV$^6$& $(1.3\pm0.2)\cdot10^{-3}$ GeV$^6$\\
\hline
\end{tabular}
}
\label{tableQ7Q8}
\end{table}
The various results for the matrix-element of $O_6^{(1)}$ are in reasonable
agreement with each other. The underlying spectral integral,
evaluated directly from data in Refs.~\citebk{NAR01},\citebk{DG2} and
 \citebk{BPQ7Q8},
or via the minimal hadronic ansatz~\citebk{KnechtQ7Q8} are in
better agreement. 
The largest source of the differences is the way the different
results for the underlying evaluation of $O_6^{(2)}$ come back into
$O_6^{(1)}$.

The results for $O_6^{(2)}$ are also in
reasonable agreement. Ref. \citebk{BPQ7Q8}
uses two approaches. First, the matrix-element for $O_6^{(2)}$ can be extracted
via a similar dispersive integral over the scalar and pseudoscalar spectral
functions. The requirements of short-distance matching for this spectral
function combined with a saturation with a few states imposes that
the nonfactorisable part is suppressed and the number and error quoted follows
from this. Extracting the coefficient of the dimension 6 operator
in the expansion of the vector and axial-vector spectral functions yields a
result comparable but with a larger error of about 0.9.
Ref. \citebk{KnechtQ7Q8} uses a derivation
based on a single resonance plus continuum
ansatz for the spectral functions and assumes a typical large $N_c$
error of 30\%. This ansatz worked well for lower moments of the spectral
functions which can be tested experimentally. Adding more resonances allows
for a broader range of results.\cite{BPQ7Q8}
Ref.~\citebk{DG2} chose to enforce all the known constraints on the
vector and axial-vector spectral functions to obtain a result. This resulted
in rather large cancellations between the various contributions making an error
analysis more difficult. A reasonable estimate lead to the value quoted.

The reason why the central value based on the same data can be so different is
that the quantity in question is sensitive to the energy regime
above 1.3~GeV where the accuracy of the data is rather low.

\section{Conclusions}
\label{conclusions}

Penguins are alive and well, they provide a sizable part of the $\Delta I=1/2$
enhancement though mainly through long distance Penguin like
topologies in the evaluation of the matrix-element of $Q_2$.
They have found a much richer use in the $CP$ violation phenomenology.
For the electroweak Penguins, calculations are in qualitative agreement
but more work is still needed to get the errors down. For the strong Penguins,
the work I have presented here shows a strong enhancement over factorisation
with $B_6$ significantly larger than one. The latter conclusion is similar
to the one derived from the older more phenomenological arguments where
the coefficients were taken at a low scale and the matrix-elements
for $Q_6$ taken from the the value of the $\Delta I=1/2$ rule.
This also indicated a rather large enhancement of the matrix-element of $Q_6$
over the naive factorisation.

\section*{Acknowledgements}
This work has been partially supported  by the Swedish Research Council
and by the European Union TMR Network
EURODAPHNE (Contract No. ERBFMX-CT98-0169).
I thank the organisers for a nice and well organised meeting and Arkady for
many discussions and providing a good reason to organise a meeting.

\end{document}